# Mécanismes fondamentaux de l'ablation laser femtoseconde en « flux intermédiaire »

*Guillaume Petite*

Laboratoire des Solides Irradiés, UMR 7642
CEA/DSM, CNRS/SPM et Ecole polytechnique
F-91128, Palaiseau

## Introduction

Qu'y a t-il de spécifique au régime femtoseconde dans l'interaction laser-solide ? On a beaucoup dit sur ce sujet, parfois sans trop réfléchir, et le propos de ce texte est de repartir des mécanismes fondamentaux, et d'un certain nombre de problèmes classiques de l'interaction laser matière, pour répondre autant que faire se peut à cette question. On constatera que certaines questions restent largement ouvertes, pas toutes spécifiques du régime femtoseconde, mais aussi que l'utilisation d'impulsions courtes ou ultracourtes (un concept en perpétuelle évolution) peut aider, par la simplification qu'elle entraîne, à apporter quelques réponses fondamentales.

Pourquoi avons nous utilisé une ligne supplémentaire dans le titre pour spécifier « en flux intermédiaire » ? Parce que nous nous plaçons dans le cadre de la physique des solides, qui a ses spécificités, et non pas dans celui de l'interaction laser-plasma (fût elle à haute densité) : nous excluons donc le cas des interactions à « haut flux », c'est à dire, du point de vue de la physique, quand les énergies d'excitations deviennent grandes devant les énergies caractéristiques de la structure électronique des solides. La dizaine d'eV est l'ordre de grandeur typique pour ce qui nous concerne, alors que l'utilisation d'éclairements très élevés (typiquement l'unité atomique) est maintenant courante, et permet d'atteindre au cours d'une impulsion, même ultracourte, des énergies beaucoup plus élevées (comme en témoigne la possibilité d'excitations en couches internes). Pour fixer les idées (mais cela dépend du problème, de la durée de l'impulsion, du matériau... ) nous nous intéressons typiquement au voisinage du seuil de claquage (au delà duquel on aboutit à la destruction du matériau), au plus quelques TW.cm$^{-2}$ en terme d'éclairement, dans le cas des diélectriques.

En s'appuyant sur la structure électronique des solides, nous décrirons d'abord les processus de base, qui contribuent à l'absorption de l'énergie – et notamment les aspects particuliers liés au fort éclairement : processus multiphotoniques, et ses modes de relaxation. Nous décrirons quelques modèles qui partant de ces processus de base, permettent une description plus globale de l'interaction. Finalement, nous évoquerons aussi les processus thermodynamiques qui peuvent résulter de l'élévation de température du matériau, qu'il s'agisse des électrons ou des ions formant le réseau. Nous verrons que ce point de vue, les impulsions femtoseconde conduisent à une situation beaucoup plus simple du point de vue physique, quant à la description d'ensemble d'une expérience d'ablation.

Nous illustrerons le plus souvent notre propos en utilisant l'exemple de matériaux isolants. Ce n'est pas parce que il n'y a pas de différences entre impulsions nanoseconde et femtoseconde du point de vue de l'interaction avec les métaux, mais c'est clairement dans le cas des isolants que les effets sont les plus importants.

## Structure des solides et absorption de l'énergie

Nous ne reviendrons pas sur les concepts de base de la physique des solides, pour lesquels le lecteur est renvoyé aux bons ouvrages du domaine, par exemple Kittel (1998) pour une approche élémentaire, Ashcroft et Mermin (2002) pour une approche un peu plus fouillée (nous donnons là les dernières éditions en Français, il en existe beaucoup d'antérieures, équivalentes à ce niveau). Nous rappelons juste quelques élément particulièrement utiles pour la suite.

### - Structure de bande et excitations électroniques dans les solides

Rappelons qu'un solide est constitué d'ions et d'électrons. Les ions sont à des positions en principe fixes (à T=0), et les électrons peuvent se répartir en deux catégories :

- les électrons « de coeur » : ils sont fortement liés aux ions du réseau, localisés. leurs énergies de liaison sont élevées (plusieurs centaines d'eV), et ils ne sont donc pas accessibles aux excitations optiques dans la gamme d'éclairement que nous considérons.

- les électrons « de valence » : les électrons les moins liés du solide. Ils sont partagés entre les ions du réseau, et sont donc plus ou moins délocalisés. Ces électrons sont ceux qui déterminent la structure atomique du solide, imposent ses symétries, bref ils sont à la base de la chimie du solide. Ils sont aussi faiblement liés, et accessibles aux excitations dans le domaine optique ou proche UV.

Nous oublions une fois pour toutes les électrons de cœur : le solide est constitué d'un arrangement périodique de cœurs ioniques plus ou moins ionisés (selon le nombre d'électrons de valence, qu'ils partagent avec les autres), et d'électrons de valence qui sont ceux qui nous intéressent. Ces électrons étant délocalisés, la représentation $\mathcal{E}(\mathbf{r})$ (énergie en fonction de la position) n'a pas d'utilité pratique – mais on peut dessiner des cartes d'isodensité à une énergie donnée – et on lui préfère la représentation en « structure de bandes » $\mathcal{E}(\mathbf{k})$, dont un exemple est donné sur la figure 1. Le cas représenté (proche du diamant) est celui d'un isolant, pour lequel le « niveau de Fermi » -dernier état potentiellement occupé en utilisant tous les électrons disponibles – a été, comme c'est l'usage pour un isolant parfait, placé au milieu de la bande interdite. $\mathbf{k}$ est le « moment cristallin » de l'électron, notion subtile dont on retiendra que $\hbar\mathbf{k}$ n'est pas la quantité de mouvement de l'électron (Ashcroft et Mermin, p. 260).

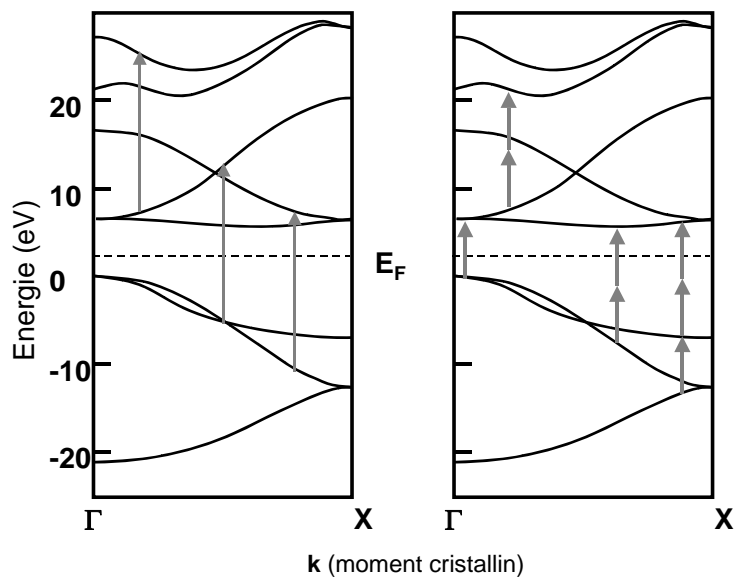

*Figure 1 : Structure de bande d'un solide cristallin parfait, et transitions électroniques possible par absorption d'un rayonnement optique. A gauche : transitions à un seul photon, a droite : transitions multiphotoniques*

On a représenté sur la figure 1 les différentes transitions possibles par absorption de rayonnement électromagnétique (e.m.). Si l'énergie des photons est supérieure à l'énergie de bande interdite, des transitions à un seul photon sont possibles (optique linéaire), mais les éclairements disponibles avec les lasers actuels autorisent bien sûr l'existence de transitions multiphotoniques. Ainsi, aucun matériau n'est réellement transparent au rayonnement d'un laser suffisamment intense !

Autre remarque concernant la figure 1 : toutes les transitions figurées sont « verticales ». Ceci résulte simplement du fait qu'un photon porte un moment négligeable en comparaison, par exemple, avec la largeur de la zone de Brillouin (ZB). Une conséquence de ceci est que ces transitions (mono- comme multiphotoniques) ne sont possibles qu'en un nombre restreint de points de la ZB : l'énergie finale des électrons excités (immédiatement après l'interaction) aura donc des valeurs spécifiques.

Remarquons aussi que nous avons fait figurer simultanément des processus d'ordres différents, et on trouve de ce point de vue dans la littérature concernant l'interaction laser-solide, de nombreuses références au phénomène d' « Above Threshold Ionisation » (ATI) dans les atomes (Agostini *et al.*, 1979). Remarquons que si une telle situation se produit dans un solide, à cause de la dispersion de la bande de valence, elle n'a aucune raison de conduire au spectre d'énergie constitué de pics régulièrement espacés de l'énergie du photon.

Nous ne commenterons pas ici l' autre mode d'excitation possible, à savoir l'effet tunnel alternatif (Agostini et Petite, 1988), non qu'il soit impossible en principe, mais il n'intervient que très marginalement aux éclairements que nous considérons ici.

Signalons, avant d'aller plus loin, que nous avons implicitement utilisé ci dessus une approximation de degré zéro, et peut être un peu moins ! En parlant par exemple d « électron excité », nous avons implicitement supposé que cet électron était indépendant des autres. Ceci est évidemment faux, et il nous aurait fallu au lieu d'électrons parler de « quasi-électrons », c'est à dire d'une particule composite, qui tient compte précisément de ces couplages que nous avons négligés, et qui permet de raisonner en particules indépendantes. C'est une question difficile qui dépasse de beaucoup le cadre de ce cours. Au delà des ouvrages généraux, le lecteur intéressé par ce problème pourront se référer à un article de revue récent (Onida, *et al.*, 2002). Quant à nous, nous persévérerons dans l'erreur, mais pas sans signaler quand même que, par exemple, calculer le spectre d'absorption optique d'un matériau sur la base de l'approximation en électrons indépendants conduit à des résultats grossièrement faux : les erreurs sur la bande interdite (la région de transparence d'un matériau isolant) peuvent facilement atteindre 100% (être sous estimée par un facteur 2). Nous ferons quand même attention à ne pas énoncer, de ce fait, de conclusions qualitativement fausses. Mais nous attirons l'attention sur le point que, particulièrement quand on consulte la littérature sur des propriétés impliquant des états excités (dans l'exemple de la figure 1 : une structure de bande de conduction), il faut s'inquiéter de savoir dans quel cadre elles ont été calculées.

### - Absorption intrinsèque, absorption de défauts

Nous avons considéré ci dessus le cas d'un matériau idéal. Ce sont évidemment des objets auxquels on a rarement affaire dès qu'on s'intéresse aux applications de l'ablation laser. Comment devons nous réviser le schéma idéal de la figure 1 dans le cas d'un matériau réel ? De quels défauts parlons nous pour commencer ?

On distingue deux types de défauts : étendus et ponctuels. La surface d'un matériau (au sens de l'interface entre un cristal parfait semi-infini et le vide) est par exemple un défaut étendu (Desjonquères et Spanjaard, 1996). Comme de plus il est périodique, on peut lui appliquer le théorème de Bloch, et définir – comme pour un cristal parfait - une structure de bande de surface. Ceci est un cas plutôt exceptionnel, et la préparation de telles surfaces requiert des moyens sophistiqués et une compétence spécifique.

Deux surfaces d'orientations différentes d'un même matériau mises en contact constituent un autre type de défaut étendu bidimensionnel : le « joint de grain ». Là encore, un joint de grain parfait possède, comme une surface une structure de bande à deux dimensions.

L'autre exemple classique de défaut étendu (1D cette fois) est la dislocation, qu'on pourrait définir de façon très générale comme une région linéaire du cristal (ligne fermée ou débouchant sur une surface ou un joint de grain) telle que (Ashcroft, 2002, p.755) :
- loin de cette région, le cristal n'est localement différent du cristal parfait que de manière négligeable.
- au voisinage de la dislocation, les positions atomiques sont fortement distordues par rapport à celles du cristal parfait.

Une dislocation n'a donc pas nécessairement une structure périodique, et le problème de sa structure électronique est beaucoup plus délicat.

L'autre grande classe de défauts est celle des défauts ponctuels. Ces défauts peuvent être intrinsèques : tout ion du réseau déplacé de sa position naturelle. Des exemples de ce type de défaut sont la lacune (un ion manque), l'interstitiel (un ion du réseau est dans une position différente de celle qu'il doit occuper dans le cristal parfait). Ces deux défauts sont souvent associés, et constituent alors une « paire de Frenkel ». Enfin, par exemple dans un alliage métallique A-B, il existe les défauts d'antisite (un atome de type A occupe une position normalement affectée à un atome de type B). Mais il existe aussi des défauts dits « extrinsèques », constitués par des impuretés soit en sire substitutionnel (exemples importants pour les lasers : le chrome ou le titane dans $Al_2O_3$), soit en site interstitiel.

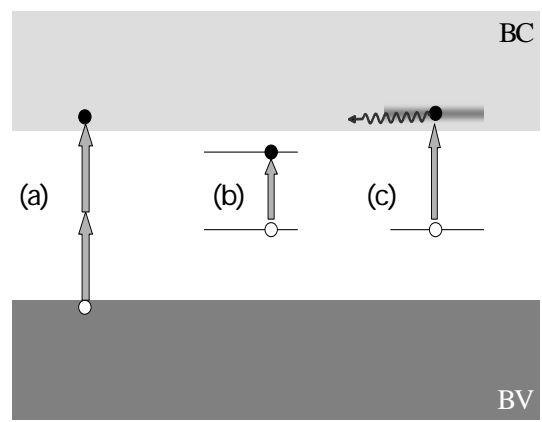

*Figure 2 : transitions interbandes (a) et transitions des défauts (b) et (c). (c) représente le cas d'une transition « ionisante » pour laquelle l'électron excité peut transiter vers les états de la bande de conduction du cristal.*

Du point de vue de la structure électronique, le fait que ces défauts soient ponctuels (donc leurs électrons de valence localisés) rend le concept de structure de bande inadéquat. C'est pourquoi sur la figure 2, nous n'avons fait figurer que les énergies autorisées (par de coordonnée **k**), ici encore pour le cas d'un isolant – cas pour lequel l'effet des défauts est déterminant – où on constate que leur présence se traduit par l'existence d'états autorisés dans la bande interdite. L'existence de tels défauts induit donc la possibilité de nouvelles transitions, d'énergie inférieure à celle des transitions interbandes.

Notons qu'il s'agit de transitions entre états de défauts, et deux cas peuvent se produire. L'état final peut, comme l'état initial, se trouver dans la bande interdite (cas b). Dans ce cas, il reste lié au défaut, et l'excitation induite donnera le plus souvent lieu à une émission de luminescence. Si au contraire l'état excité se trouve dégénéré avec les états de la bande de conduction, l'électron excité aura une forte probabilité de transiter vers ces états de conduction, et il deviendra indiscernable des autres électrons de conduction du système. Le défaut dans ce cas perd un électron et augmente donc son état d'ionisation.

Nous avons aussi fait figurer sur la figure 2 le cas (a) de transitions multiphotoniques interbandes pour illustrer le problème suivant : quelles transitions vont dominer dans le cas d'une interaction avec un laser intense. Cela dépend évidement de la densité de défauts, et des sections efficaces de transition. Pour fixer les idées, nous allons utiliser pour l'ensemble des transitions de la figure 2 une loi d'échelle moyenne, établie sur un grand nombre de cas en physique atomique (Agostini et Petite, 1988), et qui se trouve représenter aussi de manière acceptable les probabilités de transition dans un isolant (Jones *et al.*, 1989)- le cas des métaux est différent : les probabilités y sont le plus souvent plus grandes. Rappelons que la probabilité par unité de temps d'un processus multiphotonique d'ordre N peut s'écrire comme

$$W = \sigma_N F^N \tag{1}$$

où $F$ est le flux de photons incidents, et $\sigma_N$ est un coefficient appelé « section efficace généralisée » (en unité cm$^{2N}$ s$^{1-N}$ : cm$^2$ si $N=1$). C'est cette section efficace généralisée qui obéit à une loi d'échelle moyenne (qui pourra toujours être prise en défaut sur un cas spécifique) :

$$\sigma_N \approx 10^{-19} \times \left(10^{31\pm 2}\right)^{1-N} \tag{2}$$

ce qui revient à dire que quand on passe d'un ordre N au suivant, on perd 31 ordres de grandeur sur la section efficace !

Comparons sur cette base les probabilités de transitions induites dans le cas (a) et le cas (c) par une impulsion pour laquelle nous fixons la fluence $\Phi$ et laissons varier la durée $\tau$.

La probabilité (par unité de surface éclairée) d'un processus de type (a) s'écrit :

$$P_V = N_V \sigma_2 \left(\frac{\Phi}{E_P \tau}\right)^2 \tau \tag{3}$$

et pour le processus de type (c)

$$P_D = N_D \sigma_1 \left(\frac{\Phi}{E_P}\right) \tag{4}$$

où $N_V$ – resp. $N_D$ – est la densité d'électrons de valence – resp. de défauts – et $E_P$ l'énergie de photon. On remarque que si la probabilité du processus (c) ne dépend que de la fluence – normal pour un processus linéaire – celle du processus (a) dépend, à fluence fixée de la durée de l'impulsion, à laquelle elle est inversement proportionnelle. On en tire une conclusion importante, concernant l'utilisation d'impulsions femtosecondes :

> *A fluence fixée, l'utilisation d'impulsions courtes favorise les processus intrinsèques par rapport aux processus résultant de la présence de défauts*

Evidemment, l'utilisation de fluences élevées a le même résultat, mais on se souviendra que les processus de claquage (et plus généralement tous les processus impliquant une modification irréversible du matériau) obéissent à des seuil en fluence, et il y a donc une limite a cette dernière. Illustrons ce point par un calcul d'ordre de grandeur, en comparant les probabilités (3) et (4) pour deux impulsions respectivement de durée 1 ns et 100 fs. On a :

$$P_D / P_V \approx \left(N_D / N_V\right) 10^{31} \left(\frac{\tau}{N_P}\right) \tag{5}$$

où $N_P$ est le nombre de photons dans l'impulsion. Supposons une densité de défauts de l'ordre du ppm (plutôt faible). Les deux processus auront la même probabilité pour $N_P=10^{16}$ photons (énergie de l'impulsion de l'ordre de quelques mJ) pour une impulsion nanoseconde, et pour $N_P=10^{12}$ photons (énergie de l'impulsion de l'ordre de quelques fractions de µJ) pour une impulsion de 100 fs. Si la comparaison avait été faite avec un processus d'ordre 3 pour (a), on aurait abouti à des valeurs de $10^{19}$ et $10^{15}$ photons. Une conséquence de ceci est la suivante :

> *L'utilisation d'impulsions femtoseconde permet en général de travailler à des éclairements plus élevés, sans atteindre le seuil de claquage du matériau*

## Processus de relaxation de l'énergie électronique

L'électron excité par l'absorption de photons évoquée ci dessus est en interaction avec un grand nombre d'autres particules, ce qui va conduire à une relaxation plutôt efficace de l'énergie qu'il a emmagasinée. Les processus diffèrent selon l'interaction que l'on considère.

### - Relaxation radiative

L'électron est d'abord en interaction avec le trou qu'il a laissé dans la bande de valence (voir éventuellement à haute densité d'excitation, avec des trous correspondant à d'autres électrons excités). A cette interaction correspondent les processus de relaxation radiative : l'électron se recombine avec le trou en émettant un photon, dans un processus parfaitement symétrique du processus d'absorption linéaire décrit plus haut.

Signalons tout de suite qu'à l'échelle des autres processus que nous allons étudier plus bas, les processus radiatifs sont lents. Typiquement nanoseconde, quelquefois plus, rarement moins. Il ne deviennent importants que lorsque les autres processus de relaxation ont été « débranchés ». C'est par exemple le cas pour le processus de luminescence de défaut discuté à propos du cas (b) de la figure 2. Ne les négligeons pas quand même : c'est la dessus que fonctionnent par exemple tous les lasers solides.

### - Relaxation électron-électron

Nous l'avons déjà signalé, un électron excité est en interaction avec tous les autres électrons du système. Dans notre modèle en électrons indépendants, nous pouvons traiter cette interaction avec un modèle de collisions : un électron (de conduction, par exemple ) peut dans une collision avec un autre électron du système lui transmettre tout ou partie de son énergie.

Dans le cas des métaux, ce problème (qui est évidemment important du point de vue des théories du transport électronique) se traite à l'aide de la théorie du « liquide de Fermi » due à Landau (1957). Sans entrer dans les détails de cette théorie, disons simplement que la possibilité pour un électron situé au dessus du niveau de Fermi d'exciter un autre électron dépend de l'espace de phase accessible pour lui même et pour l'électron qu'il excite dans l'état final de la collision, ce dernier étant limité par le principe d'exclusion de Pauli. On en tire une dépendance du temps de vie de l'électron initial (d'énergie $\mathcal{E}_i$) qui s'exprime comme

$$\frac{1}{\tau} = a(\varepsilon_i - \varepsilon_F)^2 + b(k_B T)^2 \qquad (6)$$

où a et b sont des constantes indépendantes de $\mathcal{E}_i$ et T. Pour des électrons dont l'énergie initiale est de l'ordre de l'eV ou plus (grande devant l'énergie « thermique ») on en retiendra la dépendance en $(\mathcal{E}_i - \mathcal{E}_F)^{-2}$. L'ordre de grandeur de la durée de vie dépend évidemment de la constante $a$, mais on retiendra qu'à quelques eV pour $\mathcal{E}_i$ on aboutit à des durées de vie de quelques dizaines de femtosecondes, à plus ou moins un ordre de grandeur près.

Le cas des isolants est plus complexe. En effet, la présence d'un bande interdite implique une seuil pour l'énergie d'excitation. La question de la valeur de ce seuil n'est pas absolument claire dans la littérature. On trouve très souvent écrit que ce seuil est égal à l'énergie de bande interdite. C'est le plus souvent faux ! Le problème est illustré sur la figure 3 à partir de deux situations modèles, toutes deux ultra simplifiées : rappelons qu'il ne suffit pas de conserver l'énergie totale, mais aussi le moment total. Cela n'a pas d'incidence dans le cas où la bande de valence est sans dispersion (trous « infiniment lourds ») et on a bien là une énergie seuil égale à $E_G$. Mais on voit que dans le cas d'un matériau à « gap direct », et avec des masses équivalentes pour l'électron et le trou, ce seuil passe à $2E_G$. Mais c'est oublier de tenir compte de ce qu'à une énergie de $2E_G$, on est souvent au delà de la limite de la première ZB. Or la conservation du moment ne doit être assurée qu'à un vecteur du réseau réciproque près, ce qui complique singulièrement l'évaluation du seuil. Le lecteur est donc invité à la plus grande prudence face aux affirmations trop simples concernant l'énergie seuil dans un matériau donné. On rappellera cependant que souvent dans les oxydes, la masse des trous est très supérieure à celle des électrons, ce qui les rapproche du cas des matériaux « à bande de valence plate ».

Ces processus de collision electron-electron peuvent être représentés à l'aide de section efficaces qui peuvent soit être obtenues *ab initio*, soit à l'aide de formules empiriques. Nous donnons ci dessous la formule employée pour introduire ces phénomènes dans les calculs Monte Carlo du transport d'électrons dans les silices pour dispositifs MOS (ou ces problèmes sont critiques, puisqu'ils déterminent en partie la tenue des dispositifs) (Strobbe *et al.*, 1991)

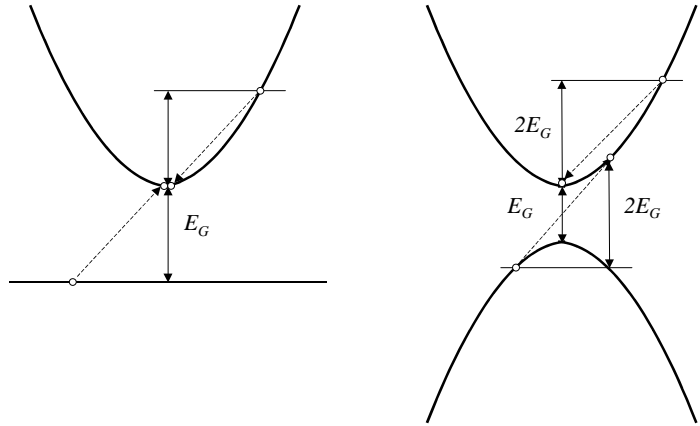

*Figure 3 : excitation de paires e-h secondaires dans un isolant. A droite : matériau à bande de valence « plate », à gauche, matériau « symétrique » (gap direct, $m_e=m_h$)*

$$f_{ee} = c\left[\frac{E/E_{th}-1}{1+D\cdot E^2/E_{th}^2}\cdot \ln\left(\frac{E}{E_{th}}\right)\right]^a \tag{7}$$

valable pour un matériau à bandes paraboliques, où $f_{ee}$ représente la fréquence des collisions électron-électron donnant lieu à une excitation de paires, $E_{th}$ l'énergie seuil, $c$, $D$ et $a$ des paramètres du modèle. On notera simplement que cette expression s'écarte fortement du modèle de type « liquide de Fermi » évoqué plus haut.

Nous n'avons envisagé jusqu'ici que les processus de collisions « individuelles ». Parmi les mécanismes de pertes d'énergie électroniques, il faut aussi tenir compte des processus collectifs. On parle de perte d'énergie par « excitation de plasmons ». Le concept de plasmon d'un gaz d'électrons libres (modèle de Drude) est bien connu : il est l'élément central des propriétés optiques des métaux. L'existence de pics caractéristiques de pertes d'énergies correspondant à l'énergie de ces plasmons est un résultat bien connu en « Spectroscopie de Pertes d'Energies Electroniques », et ceci que ce soit dans les métaux ou dans les isolants. Certes, la théorie de Drude ne peut s'appliquer aux électrons liés des isolants, mais on peut lui substituer le concept d'oscillation collective d'oscillateurs harmoniques couplés (censé représenter les oscillations des électrons de valence) pour obtenir une vision qualitative du plasmon d'un isolant. La figure 4 présente le résultat d'un calcul (dans l'approximation dite « RPA », assez bonne –pour des raisons subtiles de compensation de différents effets – dans ce type de calculs) du spectre de pertes d'énergie électronique dans le diamant (Olevano *et al.*, 2004).

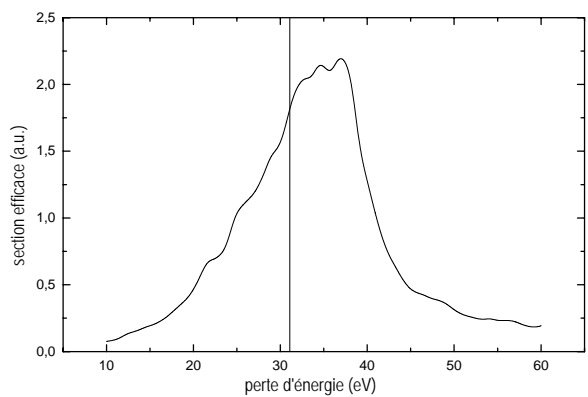

*Figure 4 : spectre de pertes d'énergie électronique dans le diamant, calculé dans l'approximation RPA. Le trait vertical à 31.5 eV montre la position du plasmon calculé avec la théorie de Drude en utilisant la densité électronique du diamant*

On constatera particulièrement deux faits sur cette figure
- la position du « pic de plasmon », la structure large située entre 25 et 40 eV, n'est finalement pas si loin de la position qu'on aurait calculée avec un modèle de Drude en utilisant la densité électronique du diamant (ceci est probablement

- du au fait que cette énergie de plasmon est grande devant « l'énergie de liaison des électrons de valence », si on assimile cette dernière à la largeur de bande interdite)
- les pertes d'énergie individuelle discutées plus haut sont présentes dans ce calcul : il s'agit du résidu à 10 eV de la section efficace. On constate que leur importance est pratiquement négligeable par rapport aux pertes par émission de plasmon (mais attention aux conditions du calcul : l'énergie de l'électron incidente est ici élevée, c.a.d. grande devant l'énergie de plasmon)

On retiendra de ceci qu'il ne faut en aucun cas, si elles peuvent intervenir (c'est à dire dès que l'énergie cinétiqude l'électron de conduction excède celle du plasmon), négliger ces pertes d'énergie d'origine collective.

### - Relaxation électron-phonon

Les électrons sont évidemment en interaction avec les ions du réseau, mais il faut évidemment tout de suite préciser que cette interaction est prise en compte dans la structure de bande du matériau. On évitera donc de parler de collision des électrons avec les ions du réseau !

Par contre, ce qui est pris en compte est seulement l'interaction avec les ions dans leur position théorique. Les vibrations du réseau impliquent des écarts à ces positions d'équilibre, et donc une modification de l'interaction avec les électrons. Un traitement perturbatif de cette interaction conduit à la notion « d'interaction électron-phonon ». Nous n' en développerons pas le formalisme ici (on peut se reporter à Petite, 1996) mais rappellerons quelques points essentiels :
- les vibrations du réseau peuvent être vues comme un ensemble d'oscillateurs harmoniques. Les « phonons » sont les quanta de vibration correspondant à ces oscillateurs.
- comme les électrons ces phonons possèdent une structure de bande (relation de dispersion $\omega(q)$ – où $\omega$ et $q$ représentent la fréquence et le moment du phonon). Selon le type d'atome et le nombre d'atomes par maille, on obtient deux types de branches correspondant aux « phonons acoustiques » et aux « phonons optiques », ces derniers seulement dans le cas où il y a deux atomes différents par maille. La forme caractéristique des ces deux branches est représentée sur la figure 5. Au voisinage du centre de la ZB on peut considérer que les phonons acoustiques représentent les oscillations de dilatation d'ensemble de la maille, les phonons optiques représentant les oscillation dipolaires des atomes différents les uns pas rapport aux autres. En limite de ZB (phonons de courte longueur d'onde), la situation est moins claire.

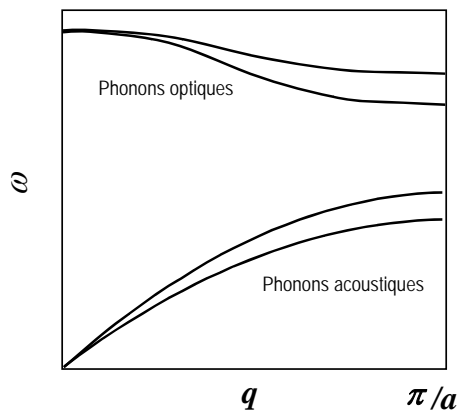

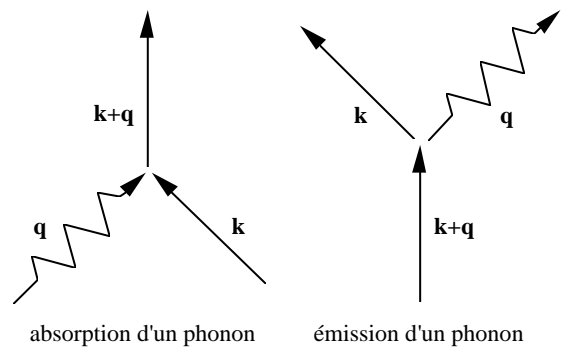

*Figure 5 : Relation de dispersion typique des phonons, montrant les deux types de branches*

*Figure 6 : Représentation schématique d'une collision électron-phonon.*

A l'aide de cette représentation, l' interaction de l'électron avec les vibrations du réseau peut être considérée comme une collision entre un électron et un phonon, au cours de laquelle le phonon est absorbé ou émis (Figure 6). On y a représenté les deux processus essentiels – et symétriques - que sont
- l'annihilation d'un électron dans l'état **k** et d'un phonon de vecteur d'onde **q**, et à la création d'un électron dans l'état **k**+**q**;
- l'annihilation d'un électron dans l'état **k**+**q**, et à la création d'un phonon de vecteur d'onde **q** et d'un électron dans l'état **k**.

Si le nombre de phonons dans le mode de vibration considéré est n, on notera que la probabilité d'émission de phonons est proportionnelle à n+1, alors que la probabilité d'absorption est proportionnelle à n (noter la similitude avec les relations d'Einstein pour l'absorption ou l'émission de photons dans un mode du champ e.m.). On notera que pour un matériau à température ambiante, en général $n \ll 1$, et que les processus d'émission spontanée de phonons vont en général dominer.

A ce stade, quelques remarques méritent d'être faites

- les phonons d'énergie la plus élevée correspondent aux modes optiques. Par exemple le quartz à deux modes de ce type, correspondant à des énergies de phonon de 63 et de 153 meV. Si on prend ce dernier, il lui correspond une fréquence de vibration de 3.6 10$^{13}$ Hz. La plupart des systèmes femtoseconde commerciaux délivrent maintenant des impulsions de l'ordre de, et souvent inférieures en durée à 50 fs. A cet échelle on peut donc considérer que, même dans le cas extrême évoqué ci dessus, les ions du réseau peuvent être considérés comme immobiles. On en tire une particularité des impulsions femtoseconde, que l'on peut exprimer ainsi :

> *A l'échelle d'une impulsion femtoseconde, on peut considérer que l'on interagit avec un réseau dont les vibrations ont été gelées.*

- Les phonons sont des particules qui transportent un moment important. Dans une collision « à trois corps » (électron-photon-phonon), l'absorption d'un photon fournit une énergie importante, mais peu de moment, alors que l'absorption (ou l'émission) d'un phonon implique peu d'énergie, mais un transfert de moment important. Cela ouvre la possibilité d'une autre classe de processus d'absorption que ceux évoqués au début de ce document : les transitions « indirectes », dont le principe est représenté sur la figure 7.

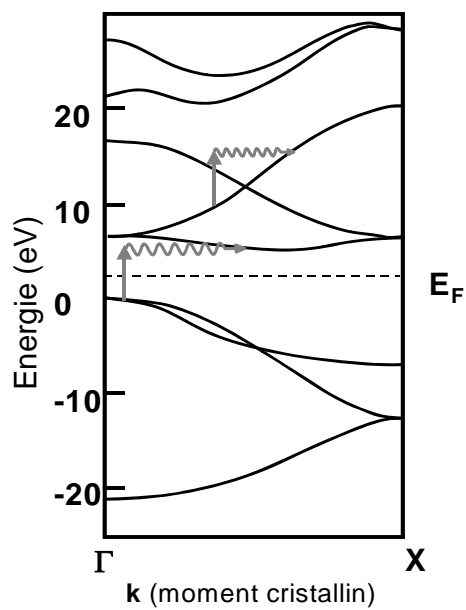

*Figure 7 : principe des transitions indirectes, impliquant l'absorption (ou l'émission) simultanée d'un photon et d'un phonon. Deux types de transitions (interbande et intrabande) ont été représentés.*

On ne cherchera pas à évaluer le rapport entre les probabilités de processus directs et indirects, car celui ci dépend à la fois de la température du matériau (par le flux de phonons) et du taux de collisions électron-phonon qui est très dépendant du matériau, ainsi que de « l'écart à la résonance » de l'état intermédiaire (en fait, beaucoup des concepts utilisés pour la physique des transitions à deux photons se transposent aisément ici). Signalons simplement que, par construction, le taux d'absorption indirecte est inférieur au taux de collision électron-phonon.

Signalons enfin, concernant le taux de collision electron-phonon, qu'il dépend de l'interaction effective entre les électrons et les ions du réseau, c'est à dire en tenant compte le cas échéant de l'écrantage de cette interaction par les autres charges du système. De ce point de vue, métaux et isolants différent profondément. L'écrantage par les autres électrons de conduction est fort dans les métaux, alors qu'il est quasiment inexistant dans les isolants. Ceci explique pourquoi le taux de collision électron-phonon est beaucoup plus grand dans ces derniers (jusqu'à des valeurs de 10$^{15}$ s$^{-1}$ par exemple dans le quartz, un ordre de grandeur caractéristique des métaux étant plus proche de 10$^{13}$ s$^{-1}$).

### - Piégeage de charges dans les isolants

Un autre propriété caractéristique des isolants est leur capacité, dans certains cas, à piéger des charges. Ce mécanisme peut être de nature extrinsèque, c'est à dire lié à la présence de défauts du matériau, ou intrinsèque, c'est à dire caractéristique de la nature même du matériau.

Le *piégeage extrinsèque* interviendra par exemple quand le matériau comporte des impuretés possédant plusieurs états d'ionisation stables, le piégeage d'un électron abaissant alors l'état d'ionisation. Ceci ne concerne pas seulement les impuretés : un autre exemple classique est la lacune d'oxygène dans $Al_2O_3$, connue sous le nom de « centre F ». Ainsi, le centre F+ (lacune une fois ionisée) peut piéger un électron pour devenir un centre F. Il existe des centres piégeant les électrons, mais aussi des centres piégeant les trous (par exemple le centre NBOHC : Non Bridging Oxygen Hole Center, dans la silice).

L'*autopiégeage* est un processus de nature très différente. C'est Landau qui le premier a émis l'idée qu'un porteur de charge pouvait spontanément se piéger dans un isolant, en remarquant que l'énergie associée à un électron dans un réseau polarisable était égale à

$$\int_V \frac{D^2}{4\pi}\left[\frac{1}{\varepsilon_\infty} - \frac{1}{\varepsilon_0}\right]dv \tag{8}$$

où $D$ est le déplacement électrique. Cette quantité étant négative, il est donc intéressant du point de vue énergétique de localiser une charge dans un isolant. Plus tard la théorie du « polaron » a fait l'objet d'un nombre considérable de travaux variés (Frölich 1954 ; Feynmann, 1955 ; Austin et Mott, 1969).

Bien sûr, cette localisation, du point de vue de la structure électronique a aussi un coût, dont l'ordre de grandeur est donné par l'énergie de saut d'un site du réseau au site voisin, c'est à dire la largeur de la bande de conduction. C'est donc l'équilibre entre ces deux quantités qui va déterminer si un matériau est apte ou non à piéger tel ou tel type de charge (électron, trou, ou « exciton », c.a.d. une paire e-h en interaction). Toyozawa (1980) a pu établir théoriquement une classification des matériaux – isolants et semiconducteurs - selon le type de charges qui s'y autopiégent qui est en bon accord avec les observations, et à propos de laquelle il est important de remarquer que ce sont les propriétés acoustiques et élastiques des matériaux qui apparaissent comme le facteur déterminant.

L' « Exciton auto-piégé » (STE pour « self trapped exciton ») a été l'objet d'études particulièrement poussées (Song et Williams, 1993), car on a pu montrer que l'existence dans un matériau de STE impliquait en général une forte sensibilité aux radiations ionisantes. La localisation d'un électron et d'un trou sur des sites voisins du réseau cristallin donne naissance à une distorsion du réseau qui stabilise le STE. Le STE est un intermédiaire entre l'excitation électronique et la création de défauts lacunaires. La figure 8 Illustre ce mécanisme en montrant d'une part la relaxation de réseau associée au STE du quartz (Fisher *et al.*, 1990), et la structure électronique de cet objet (Schluger, 1988), montrant bien son caractère métastable (sa recombinaison radiative étant interdite par le caractère singulet-triplet de la transition).

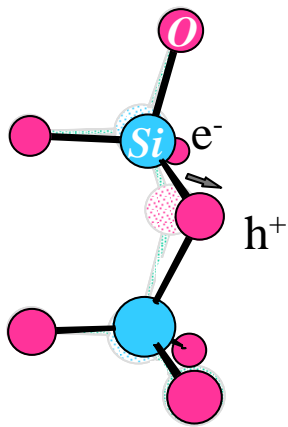
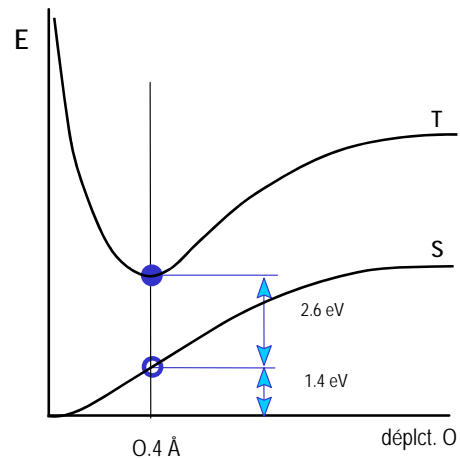

*Figure 8 (a): STE d'u quartz. Le piégeage d'un électron sur un silicium et d'un trou sur l'oxygène voisin provoque un déplacement de 0.4 Å de l'oxygène, dans la direction de l'axe c*

*Figure 8 (b) : Structure électronique du STE du quartz. Au déplacement de 0.4 Å de l'oxygène correspond un minimum d'énergie du premier état excité, qui est un état triplet*

Notons que de nombreuses expériences ont démontré l'existence d'une luminescence associée au STE du quartz, à une énergie de photon de 2.8 eV, en excellent accord avec le calcul de Schluger (1988). Par ailleurs notons que cette recombinaison s'accompagne du retour à sa position d'équilibre de l'ion oxygène, et donc de la libération dans le réseau d'une énergie élastique de 1.4 eV par STE. Un autre effet notable du STE est qu'il s'accompagne d'une augmentation de volume de la maille qui peut atteindre plusieurs dizaines de pour cent, ce qui illustre bien que l'existence d'un STE fortement lié est associée à un fort couplage avec les phonons acoustiques. Une autre classe importante de matériaux dans lesquels les STE jouent un rôle important sont les halogénures alcalins.

# Aspects cinétiques dans les modèles d'interaction laser solide

Avec les briques élémentaires que nous avons introduites ci dessus, il faut maintenant construire un modèle global de l'interaction. Puisque nous nous intéressons ici au régime femtoseconde, nous porterons une attention particulière aux aspects cinétiques de ces modèles. Il est bon de souligner ici que la plupart des modèles que nous évoquerons ne sont pas utiles que dans le domaine de l'ablation laser et que, dans de nombreux cas, ils ont été développés pour décrire d'autres situations.

### - Apport d'autres expériences d'irradiation

L'excitation optique d'un matériau se caractérise par le fait que seuls les électrons du système sont concernés par l'interaction initiale, avec néanmoins comme résultat final des modifications plus où moins massives de la structure atomique du matériau (formation de défauts, changements de phase, ablation....).

D'autres types d'irradiation procèdent aussi par excitation électronique, avec des résultats comparables. C'est notamment le cas des irradiations aux ions lourds rapides. Ces dernières partagent avec les lasers femtoseconde la caractéristique de déposer l'énergie dans un temps court devant le temps de réaction du réseau (on est même ici plus proche de $10^{-17}$ s que de $10^{-13}$ !).

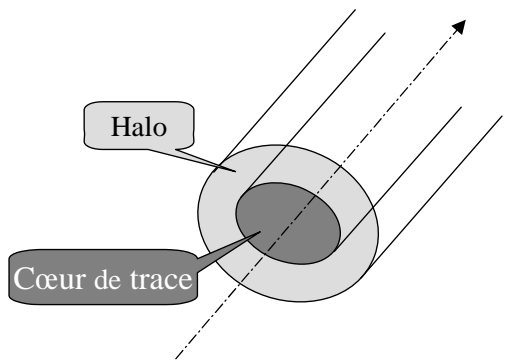

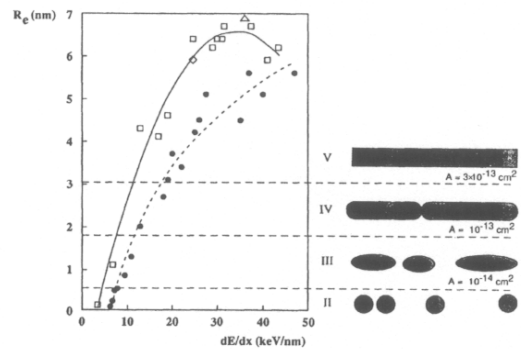

*Figure 9 a : Géométrie d'une trace d'ion lourd dans un matériau sensible au « pouvoir d'arrêt électronique », avec un coeur de trace siège d'une transformation de phase (ex : amorphisation) et un halo constitué par une zone riche en défauts ponctuels*

*Figre 9 b : évolution de la structure d'une trace dans le YIG en fonction de l'énergie déposée par unité de longueur, avec un seuil, suivi de l'apparition de zone endommagées, qui forment au forts dépôts d'énergie une trace continue . Les deux courbes illustrent l'effet de la vitesse de l'ion sur le rayon de la trace (à énergie déposée constante)(Meftah, 1993)*

On retiendra qu'un ion lourd rapide (dont la vitesse est proche de celle des électrons de cœur des ions du matériau) peut amorphiser localement ce dernier, même si seuls ses électrons sont au départ concernés par l'interaction. La structure de la trace amorphe ainsi obtenue est complexe : elle se compose d'un cœur amorphe entouré d'un « halo » qui est une zone riche en défauts ponctuels. La trace, en fonction du taux de dépôt d' énergie linéique (TEL) apparaît au dela d'un certain seuil (dépendant du matériau) et peut avoir une structure discontinue continue (bas TEL) ou continue (hauts TEL) (Meftah *et al*, 1993).

Tous les matériaux ne sont pas sensibles de manière identique au dépôt d'énergie électronique. Certains isolants ($SiO_2$, grenats, halogénures alcalins..) ont des seuils d'amorphisation bas (qques keV/nm) alors que d'autres au contraire sont pratiquement insensibles ($Al_2O_3$ et MgO sont parmi les matériaux les plus résistants). Les métaux ont très longtemps été considérés comme insensibles au dépôt d'énergie électronique (ce qu'on interprétait comme un résultat de l'écrantage quasi total des interactions électroniques), mais on a observé dans certains métaux et alliages intermétalliques l'apparition de transformations de phase locales (Dammak *et al*, 1993), telles que par exemple l'apparition de phases haute pression du titane. Il est donc maintenant admis que de nombreux métaux (mais jamais, par exemple les métaux nobles) sont sensibles à une excitation électronique intense et instantanée.

Trois modèles ont été proposés pour expliquer dans les différents systèmes cette apparition de défauts : le modèle de la « pointe thermique », fondé sur le modèle « à deux températures » de Kaganov *et al*.(1956) – et il est intéressant de constater ici que les utilisateurs de ces modèles se sont basés, pour les exploiter, sur des sonnées obtenues par des expériences d'interaction laser-solide en régime femtoseconde (Fujimoto *et al*, 1984 ou Brorson *et al*, 1990) – un modèle d'explosion coulombienne (Fleisher *et al*, 1975 et Dunlop *et al*, 1992) et, pour les isolants, un modèle fondé sur les effets excitoniques (Itoh, 1989). On retrouve - mais c'est naturel tant les domaines sont proches par la physique, même si les outils sont très différents – les grands modèles utilisés pour l'ablation laser.

Un autre type d'expérience très utiles pour valider les calculs de type Monte Carlo (Fitting, 2004), qui permettent de traiter de façon détaillée une séquence de processus de relaxation tels que ceux décrits plus haut, en y incluant éventuellement les processus de piégeage-dépégeage, sont les expériences en Microscopie Electronique à Balayage. On dispose en effet de conditions bien définies pour le faisceau incident, et d'une multiplicité d'observables (rendement d'émission secondaire, Cathodo-Luminescence...... ) qui permettent de caractériser le transport électronique et les modifications du matériau (formation de défauts - Fitting *et al.*, 2003 -, piégeage des charges – Meyza *et al.* , 2003 -....)

### - Interlude : le "Lambropoulos doom"

Arrêtons nous un instant sur un point essentiel, dont la paternité est attribuée au physicien Lambropoulos pour son illustration dans le cas ultra-simple des interactions laser-atome (d'où le nom de « Lambropoulos doom », ou « sort de Lambropoulos », en Français), mais qui s'applique à toute interaction entre un laser et un système physique. Illustrons le sur le cas de l'ionisation multiphotonique d'un atome (Figure 10). Mis en présence d'un champ laser intense, quelle que soit la longueur d'onde de celui ci, l'atome s'ionise du fait de l'absorption multiphotonique. L'interaction se développe dans le temps (dans les limites de la durée de l'impulsion), et si la probabilité totale d'ionisation atteint l'unité avant le maximum de l'impulsion, l'atome ne verra jamais le champ laser maximum (ce dernier étant le plus souvent utilisé par le malheureux expérimentateur pour définir son expérience, ce qui ne pouvait échapper à un théoricien affûté !)

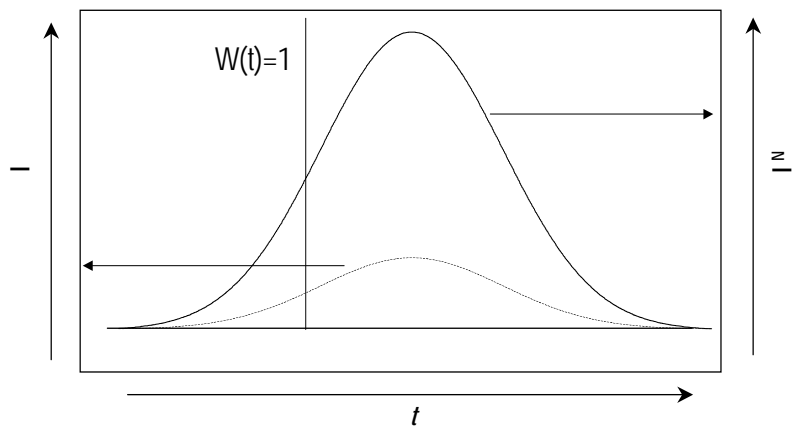

*Figure 10 : Illustration du « Lambropoulos doom » dans le cas d' l'ionisation multiphotonique d'atomes. Ligne interrompue : éclairement laser, ligne continue : probabilité d'ionisation par unité de temps. W(t) : probabilité totale d'ionisation.*

Ce comportement vaut évidemment pour tout système qui réagit à l'énergie qu'on lui injecte. La seule contre-mesure consiste alors à injecter l'énergie plus rapidement que le temps caractéristique d'évolution physique du système (temps d'ionisation, ci dessus, plus bas, nous considèrerons l'évolution thermodynamique d'un solide...) de façon à rendre son évolution négligeable pendant la durée de l'impulsion. On en tire la règle

> *L'utilisation d'impulsions les plus courtes possibles permet, dès qu'on a affaire à un système dont les propriétés évoluent sous l'influence de l'énergie injectée, de supprimer, ou au moins de minimiser les variations de propriétés du système pendant l'interaction*

Un excellent exemple, quoique déjà ancien, de cette problématique dans le cas des interactions laser-solide peut être trouvé dans le travail de Bechtel *et al.* (1975, 1977) sur l'effet photoélectrique multiphotonique.

Mais connaître les évènements élémentaires qui interviennent dans l'interaction laser-solide n'est pas assez pour décrire une expérience d'interaction, par exemple d'ablation laser. Encore faut-il pouvoir les organiser dans un modèle « global » de l'expérience en question. De quoi disposons nous ?

### - Relaxation électron-phonon dans les métaux : le modèle « à deux températures »

Plutôt que de considérer les processus collisionnels individuels dans les métaux, le modèle le plus couramment appliqué à la physique des interactions laser-métaux considère les électrons d'une part et le réseau d'autre part comme deux systèmes thermodynamiques à l'équilibre interne – notion qui a de quoi faire frémir un thermodynamicien : qu'elle soit de Boltzmann ou de Fermi-Dirac, la température décrivant la distribution est toujours celle du thermostat - que l'on peut décrire par une température électronique ($T_e$) et une température ionique ($T_i$), température du réseau. Le couplage électron-phonon intervient alors simplement comme un paramètre phénoménologique déterminant les échanges d'énergie entre ces deux systèmes. Les températures électroniques et ioniques obéissent au système d'équation

$$C_e \frac{\partial T_e}{\partial t} = \nabla(K_e \nabla T_e) - g(T_e - T_i) + A(\mathbf{r},t) \qquad (9)$$

$$C_i \frac{\partial T_i}{\partial t} = \nabla(K_i \nabla T_i) + g(T_e - T_i)$$

où $C_e$ et $C_i$ sont respectivement les capacités calorifiques électronique et ionique, $K_e$ et $K_i$ les conductivités thermiques et $g$ une constante caractérisant le couplage électron-réseau. Le laser n'intervient plus ici que par un terme source de chaleur $A(\mathbf{r},t)$ dans l'équation régissant $T_e$. On néglige parfois le terme de conductivité dans l'équation sur $T_i$, car cette diffusion est lente. On néglige aussi (sans le dire !) dans l'équation de la température ionique les pertes radiatives éventuelles (un terme qui serait symétrique du terme source $A(\mathbf{r},t)$).

On ne trouve pas facilement une solution analytique à ce problème qui doit, le cas échéant, être résolu numériquement. On peut toutefois faire un certain nombre de remarques sur les échelles de temps impliquées dans notre problème

Il y a plusieurs déterminations expérimentales du temps de relaxation électron phonon (nous entendons par là le temps de mise à l'équilibre entre les électrons et le réseau). Ces temps sont de l'ordre de la picoseconde ou moins. A l'échelle d'une impulsion nanoseconde, on peut donc considérer que la température électronique est toujours à l'équilibre avec celle du réseau, et le système d'équation ci-dessus peut être ramené à une équation pour le réseau avec terme source. Nous sommes alors ramenés au problème du transport de la chaleur : les effets du laser sur le matériau sont d'origine thermique (fusion, sublimation). *A contrario*, on considérera une impulsion subpicoseconde telle qu'on sait les obtenir actuellement (qqs dizaines de femtosecondes) comme une excitation instantanée du gaz d'électrons. Ces différences de comportement sont illustrées sur la figure 11 où on a représenté les deux cas extrêmes discutés ci dessus. Cette figure n'a qu'une ambition qualitative (il faudrait connaître le détail des propriétés électroniques et thermiques du matériau pour faire plus), mais montre la différence essentielle des comportements en impulsion courte et longue.

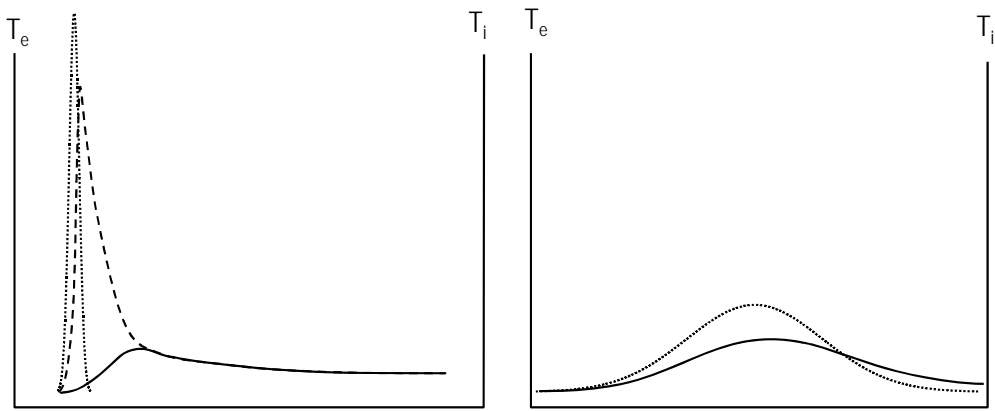

*Figure 11 : Deux cas opposés d'évolution thermique d'un matériau soumis (à gauche) à une impulsion ultracourte (à droite) à une impulsion longue. En pointillé : impulsion laser, tirets : température électronique, traits pleins : température de réseau.*

Considérons les deux cas : pour l'impulsion ultra-courte, on observe d'abord (pendant la durée de l'impulsion) une phase de « chauffage électronique ». Le réseau est à peine perturbé pendant cette phase. Suit une phase de refroidissement électronique, pendant laquelle au contraire la température de réseau augmente, jusqu'à atteindre l'équilibre électrons-réseau, la suite étant gouvernée par la diffusion thermique classique.

Dans le cas – à droite - de l'impulsion longue (c.a.d. par rapport à la durée de la phase de « refroidissement électronique » évoquée ci dessus), température électronique et température de réseau évoluent de concert. On a fait figurer un léger retard de la température par rapport à la cinétique de l'impulsion laser pour rester général, car celle ci peut être longue du point de vue « électronique » sans l'être par rapport au cinétiques de diffusion thermique.

Sans prétendre être exact de ce point de vue, nous avons sur la figure 11 illustré une situation où les fluences portées par les deux impulsions sont comparables (mais le rapport des chaleurs spécifiques électrons/réseau n'est pas réaliste). On constate qu'une conséquence en est que les températures de réseau atteintes dans les deux cas sont comparables. On retiendra la règle :

> *Le fait d'utiliser des impulsions ultracourtes n'a aucune raison d'impliquer qu'il ne peut y avoir d'effets thermiques.*

Par contre on constate aussi que dans le cas de l'impulsion ultra-courte, pendant la phase transitoire, la température électronique peut être supérieure de plusieurs ordres de grandeur à la température de réseau. Par conséquent :

> *Les phénomènes très dépendant de la température électronique peuvent être, à fluence équivalente, fortement amplifiés par l'utilisation d'impulsions ultra-courtes*

Ce problème de la relaxation dans les métaux a fait l'objet de plusieurs expériences utilisant des techniques « pompe/sonde » en impulsion ultracourtes, afin de mesurer les temps de relaxation électron-phonon. Deux méthodes ont essentiellement été utilisées : la réflexion transitoire (Schoenlein *et al.*, 1987 ; Groeneveld *et al.*, 1995), et la spectrométrie de photoémission en mode pompe-sonde (Fann *et al.*, 1992a, 1992b ; Schuttenmaer *et al.*, 1994, Hertel *et al.*, 1996). Cette dernière méthode permet en particulier d'observer l'évolution de la fonction de distribution électronique du métal excité par l'impulsion « pompe » avec des résolutions temporelles sub-picoseconde. Le modèle à deux températures est fondé sur l'idée que le temps de thermalisation des électrons entre eux (due aux interactions *e-e*, supposé être de l'ordre de la période plasmon, $10^{-15}$ s ou moins) est beaucoup plus court que le temps de thermalisation de ceux-ci avec le réseau (collisions *e-ph*, de l'ordre de la picoseconde). L'expérience de Fann *et al.* (1992) montre que rien n'est moins sûr : les distributions mesurées ont, jusqu'à des temps de l'ordre de la picoseconde, un fort caractère non-thermique.

Un modèle basé sur l'équation de Boltzmann et tenant compte de ce caractère non-thermique de la distribution électronique, en incluant les deux termes (électron-électron et électron-phonon) dans l'intégrale de collision est proposé dans Groeneveld *et al.* (1995) et rend bien compte de l'expérience. On se rapproche là des techniques de calcul que nous discuterons dans le cas des isolants. Rethfeld *et al.* (2000) on par ailleurs montré, en utilisant aussi un modèle basé sur l'équation de Boltzmann que les résultats obtenus avec un modèle à deux températures sont valide à l'approche du seuil de dommage et au delà, mais qu'une description plus fine de la distribution électronique (qui dépend du temps) est absolument nécessaire en deçà, ce qui pourrait être (vu les conditions très différentes des expériences) une explication des différences entre les résultats de Scoenlein *et al.* (1987) et de Fann *et al.* (1992).

Signalons enfin que nous ne connaissons pas d'exemple d'application du modèle d'explosion coulombienne aux interactions laser-métaux. On comprend bien cependant que ce qui relève en fait, dans le cas des ions lourds rapides, d'un effet de sillage de l'ion, n'a aucune raison de s'appliquer ici, en tout cas aux éclairements que nous considérons (les effets de sillage laser existent, mais à haut flux seulement).

### - Modèles cinétiques dans les interactions laser-isolants

Le cas de l'interaction laser-isolant est beaucoup plus difficile à traiter avec des modèles de ce type. Le concept de température électronique n'est plus directement connecté (à cause de la présence d'une bande interdite) à l'absorption d'énergie, cette dernière devenant d'ailleurs très non linéaire, de sorte que le terme source *A* est difficile à modéliser.

La compréhension des phénomènes à l'origine du claquage optique de matériaux isolants, et de comment ils dépendent de la durée de l'impulsion, a mobilisé un grand nombre de chercheurs. Dans le cas de $SiO_2$, on a bénéficié aussi de son importance pour l'électronique MOS, ce qui a provoqué un intense travail de modélisation, à l'aide de la méthode Monte Carlo, du transport électronique dans ce matériau (Fischetti *et al.*, 1985), étendu plus tard au cas des interactions laser-solide (Arnold et Cartier, 1992).

L'étude du seuil de claquage de matériaux optiques, et d'abord de la silice, en fonction de la durée de l'impulsion a été menée par plusieurs auteurs (Jones *et al.*, 1989, Lenzner *et al.*, 1998, Tien *et al.*, 1999, Stuart *et al.*, 1996), avec des résultats parfois contradictoires (expériences mono- ou multipulse, méthode de détermination du seuil...) mais un comportement général émerge (voir figure 12): aux grande durées d'impulsion, la fluence seuil évolue comme $\tau^{1/2}$, ce qui s'explique bien en supposant que toute l'énergie injectée dans les électrons de conduction est répartie sur les phonons, et que le seuil est alors déterminé par la diffusion thermique. ce comportement cesse d'être valide pour des durées inférieures à quelques picosecondes, et la question du mécanisme se pose.

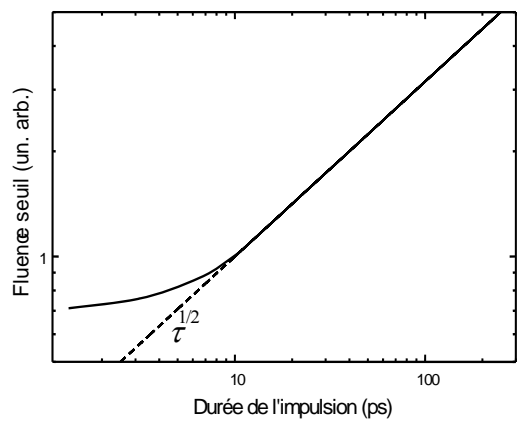

*Figure 12 : évolution caractéristique du seuil de claquage d'un isolant (ex : $SiO_2$) en fonction de la durée d'impulsion*

Plusieurs modèles ont été développés pour comprendre ce problème. On peut les classer en deux catégories : des modèles de type « cinétique chimique » qui ne considèrent que les populations, des modèles basés sur l'équation de Boltzmann, qui prennent en compte de façon détaillée les distributions d'énergie des électrons.

Les ingrédients physiques introduits dans ces modèles sont les mêmes, et ils ont été décrits plus haut. Ils comprennent :
- deux mécanismes d'absorption d'énergie laser par les électrons : absorption multiphotonique interbande, et chauffage des électrons de conduction par le mécanisme d'absorption indirecte (électron-photon-phonon) dans la bande de conduction (on parle de « Joule heating »). La possibilité d'absorption multiphotonique directe dans la bande de conduction n'est pas incluse.
- deux mécanismes de relaxation : l' « ionisation par impact » (relaxation e-e par excitation de paires e-h individuelles) et relaxation électron-phonon (les pertes par excitation de plasmon sont négligées).

Les deux types de modèles, dans les implémentations qui ont été proposées, donnent des conclusions différentes, voire pratiquement opposées.

Stuart *et al.* (1996) ont proposé un modèle de cinétique basé sur une équation de Fokker-Planck, utilisant pour données d'entrée une énergie de seuil pour l'excitation de paires e-e égale à l'énergie de bande interdite, et les résultats d'Arnold et Cartier (1992) pour le couplage électron-phonon. Les conclusions de ce modèle sont en particulier que la distribution d'énergie des électrons est stationnaire (seule la densité croît), ce qui permet de le ramener en fait a des équations de taux, et que l'injection d'électrons dans la bande de conduction est dominée par le mécanisme d'avalanche électronique, l'injection multiphotonique à partir de la bande de valence ne fournissant qu'une faible fraction initiale des électrons de conduction (voir figure 13).

Des modèles plus récents (Kaiser *et al.*,2000), utilisant les mêmes ingrédients de base, mais avec un formalisme utilisant l'équation de Boltzmann d'une part, et prenant en compte la conservation du moment dans la détermination de l'énergie seuil (ainsi que l'énergie d'oscillation de l'électron dans le champ e.m.) arrivent à des conclusions radicalement différentes : la distribution énergétique des électrons dépend fortement du temps d'une part, et que d'autre part il y a un délai appréciable (qui dépend de la fluence, mais qui est typiquement de plusieurs centaines de femtosecondes au voisinage du seuil) avant que le mécanisme d'avalanche n'apporte une contribution significative. Une proposition récente (Rethfeld, 2004), en utilisant une dicrétisation des niveaux d'énergie électronique, permet de simplifier suffisamment le formalisme tout en reproduisant très correctement les prédiction du modèle complet. Cela ouvre la possibilité d' introduire dans la modélisation toute la complexité des mécanismes fondamentaux, dont certains sont encore actuellement négligés. Ce modéle permet en particulier de comparer les résultat d'une approche « cinétique chimique » (basée sur des équation de taux, en essence le modèle de Stuart et al., 1996) avec une approche conservant la variation temporelle de la distribution, ce qui donne des résultats très différents (voir figure 14).

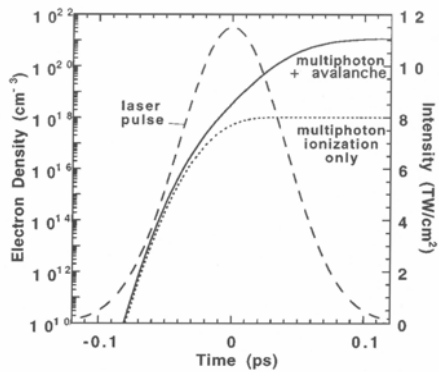
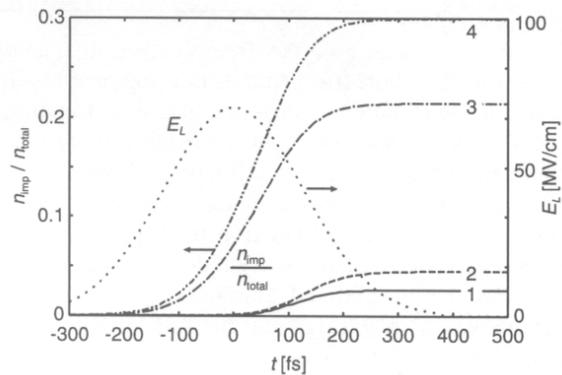

*Figure 13 : évolution de la densité électronique dans la silice soumise à une impulsion laser IR (1064 nm), montrant les contributions respectives de l'injection multiphotonique et du mécanisme d'avalanche électronique, d'après Stuart et al. (1996)*

*Figure 14 : comparaison, d'après Rethfeld (2004) des contributions de l'ionisation par impact avec un modèle « cinétique chimique » (3,4) et un modèle conservant l'évolution temporelle de la distribution d'énergie des électrons de conduction (1,2).*

Cela dit, tous ces modèles reproduisent correctement le comportement de la figure 12, ce qui montre simplement que le seuil de claquage (quelle que soit la définition qu'on en donne) n'est pas une observable suffisante pour décider de la validité de telle ou telle approche. Il faut se rapprocher à la fois des quantités physiques de base importantes, et aussi utiliser la résolution temporelle des impulsions femtoseconde pour avoir une vision « résolue en temps » du problème.

# Expériences résolues en temps sur le problème du claquage optique

Une première idée consiste à étudier le seuil de claquage provoqué non pas par une impulsion, mais par deux impulsions femtosecondes décalées dans le temps (Li *et al.*, 1999). La première impulsion est fixée 30% en dessous du seuil de claquage en impulsion unique (SP), et a pour rôle de fournir les électrons de conduction que la deuxième impulsion va « chauffer », pour provoquer le mécanisme d'avalanche prévu par Stuart *et al.* (1996). Cette mesure, effectuée aussi bien sur la silice fondue (FS) que sur un borosilicate (BBS) a donné un résultat inattendu (voir figure 15) : en fonction du délai entre les deux impulsions (jusqu'à environ 200 fs) le seuil de claquage pour le couple d'impulsions (DP) augmente, mais tout en restant en dessous du seuil correspondant à une impulsion unique. Par contre, au forts délais (nanoseconde, voir figure 16) le seuil DP rejoint la valeur SP.

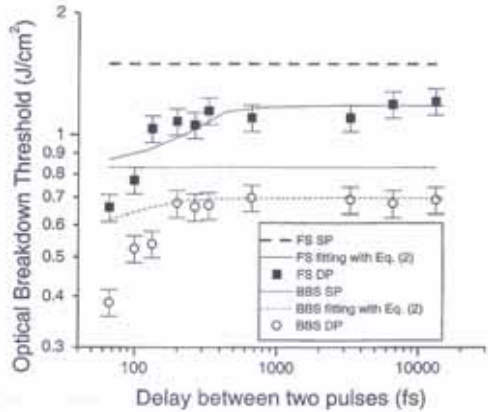
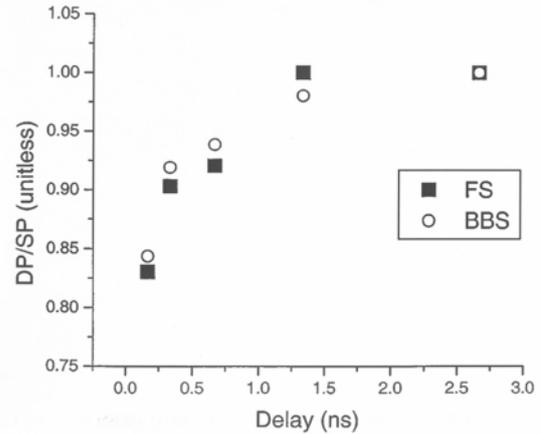

*Figure 15 : évolution du seuil de claquage pour un couple d'impulsions (DP) aux courts délais (Li et al., 1999)*

*Figure 16 : Evolution du rapport entre le seuil de claquage à deux impulsions sur le seuil en impulsion unique au forts délais (idem)*

FS : Silice fondue, BBS : Borosilicate, SP : impulsion unique, DP : double impulsion

Pour simuler ces résultats, les auteurs ont introduit une équation de taux incluant un terme de perte sur la densité d'électrons qui reproduit qualitativement l'évolution observée (voir les courbes ajustées sur la figure 15). Il se trouve que l'origine de ce comportement est connue (Petite *et al.* 1999), grâce à des expériences de mesure pompe-sonde directes de la densité d'électrons libres injectées. Ces mesures sont fondées sur les modifications de la constante diélectrique du matériau transparent, induites par les porteurs injectés. Ces modifications peuvent être observées par une mesure interférométrique de l'indice instantané du matériau. On peut montrer que cet indice s'exprime (en supposant, ce qui est le cas dans l'expérience que les modifications sont suffisamment faibles pour justifier un développement au premier ordre) comme suit :

$$n(r,t) = n_0 - \frac{1}{2n_0} \frac{N_{bc}(r,t) e^2}{m_e \varepsilon_0 \omega^2} + \frac{1}{2n_0} \sum_p \left( \frac{N_p(r,t) e^2}{m_p \varepsilon_0} \frac{1}{\omega_p^2 - \omega^2} \right) \qquad (10)$$

où on été introduits non seulement les porteurs libres (densité $N_{bc}$, masse effective $m_e$) mais aussi les porteurs piégés dans la bande interdite (densité $N_p$, profondeur du piège $\omega_p$, masse effective $m_p$, $\omega$ représentant la pulsation laser). On voit que les électrons libres abaissent l'indice (c'est pour cela que l'indice d'un plasma est inférieur à 1), alors que les électrons piégés l'augmentent, quand les pièges sont suffisamment profonds dans la bande interdite. On a pu ainsi mesurer, d'abord dans le quartz (Audebert *et al.* 1994), puis dans d'autres oxydes (Guizard *et al.*, 1995) l'évolution de la densité de porteurs libres injectés, avec une résolution temporelle de l'ordre de 60 fs. Le résultat ces mesures est montré sur la figure 17, pour $SiO_2$, $Al_2O_3$ et $MgO$.

Considérons le résultat sur $SiO_2$, le pic positif initial correspond à un effet qui n'est pas inclus dans l'eq (10). Il s'agit de l'effet Kerr, par lequel l'éclairement de la pompe provoque une augmentation de l'indice à la fréquence sonde. ce pic caractérise donc la présence simultanée de la pompe et de la sonde dans le matériau : il fixe l'origine des délais, et sa forme définit notre résolution temporelle. Le reste du comportement est conforme aux prévisions de l'eq (10) : la baisse de l'indice (déphasage négatif) signe la présence d'électrons de conduction dans le matériau, avec une densité qui relaxe avec un temps caractéristique de 150 fs, en excellent accord avec les observations de Li *et al.* (1999). A un délai proche 500 fs, le déphasage change de signe, ce qui signifie que les porteurs sont maintenant majoritairement piégés sur des sites profonds dans la bande interdite. Ce piège est stable pendant plusieurs picoseconde (en fait des mesures ultérieures ont montré qu c'était le cas jusqu'à plusieurs dizaines de picosecondes au moins.

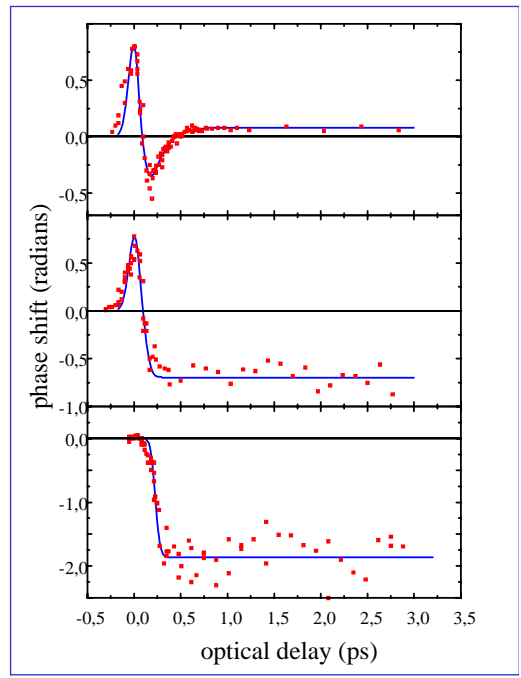

*Figure 17 : Evolution de l'indice d'un matériau dans lequel une impulsion femtoseconde intense a injecté une densité de porteurs dans la bande de conduction. De haut en bas : $SiO_2$, $Al_2O_3$ et $MgO$. Le pic positif au délai zéro est du à l'effet Kerr. Un indice abaissé (déphasage négatif) est le signe de la présence de porteurs libres.*

Sur la base de cette étude directe de la densité de porteurs injectés, on comprend. bien le résultat de Li *et al.* (1999). Quand les deux impulsions sont présentes ensemble, le seuil de claquage est évidemment abaissé. La cinétique d'augmentation qui suit résulte du piégeage des porteurs dans la bande interdite. Le seuil reste inférieur au seuil SP, car ces porteurs piégés sont plus faciles à réexciter que les électrons de valence (ils constituent un réservoir supplémentaire d'électrons facilement accessibles). Quant à la nature des pièges, il s'agit dans le cas de $SiO_2$ du STE dont nous avons parlé plus haut , et le comportement à long terme du seuil de claquage DP peut s'expliquer simplement sur la base des différents modes de relaxation du STE, essentiellement par recombinaison non radiative, mais aussi par la formation de centres colorés (centres E′ dans la silice, F dans les halogénures alcalins – Martin *et al.*, 1997)

Notons que le comportement de $SiO_2$ est particulier. On voit en effet que la durée de vie des électrons de conduction est beaucoup plus importante dans $Al_2O_3$ et dans $MgO$ (en fait de l'ordre de 100ps). Cela signifie que dans ces matériaux les électrons ne se piègent pas, ou que les pièges sont peu profonds et ne peuvent être distingués des électrons de conduction à la fréquence sonde (ici, l'harmonique 2 du laser Titane-Saphir).

Il est possible d'appliquer la même technique de mesure, en fonctionnant non pas à délai variable et à éclairement de la pompe fixé, mais au contraire à délai fixe, et en faisant varier l'éclairement de la pompe, ce qui devrait apporter des renseignements précieux sur le mécanismes d'injection. Le résultat de deux mesures de ce type (Quéré *et al.* , 1999) sur $SiO_2$ et $MgO$ est représenté sur la figure 18. On constate un comportement globalement semblable : d'abord une forte croissance, puis une saturation nette du signal. cette dernière est due au fait qu'au fortes densités électroniques atteintes (qqs $10^{19}$ à $10^{20}$ $cm^{-3}$) l'absorption par les porteurs libres devient si forte que le faisceau pompe ne pénètre plus dans l'échantillon (et même si la densité en surface continue à croître, la sonde, qui moyenne sur un certain volume, ne peut la mesurer correctement). Ceci nous conduit à insister sur le fait que dans de telles situations, une modélisation précise de la propagation du faisceau est essentielle.

Dans $MgO$, on constate que jusqu'au seuil de claquage, le comportement de la densité de porteurs en fonction de l'éclairement laser est typiquement multiphotonique. Dans $SiO_2$, le seuil (qui est plus élevé) est nettement situé dans la saturation qui démarre à des densités plus faibles (très probablement parce que la section efficace d'absorption, liée au couplage électron-phonon, est plus élevée). Par ailleurs, dans la partie de croissance rapide, on n'observe pas une loi multiphotonique mais plutôt une loi de type « effet tunnel alternatif », que nous n'avons pas discuté ici, mais qui est un phénomène prenant le relais de l'absorption multiphotonique aux hautes intensités. En tout cas, on n'observe aucun indice d'un comportement de type « avalanche » (qui donnerait une croissance quasi-exponentielle de la densité), et ceci à été confirmé par des expériences utilisant des impulsions plus longues. Il semble donc que ces expériences confirment plutôt les prédictions de Kaiser *et al* (2000) et de Rethfeld (2004).

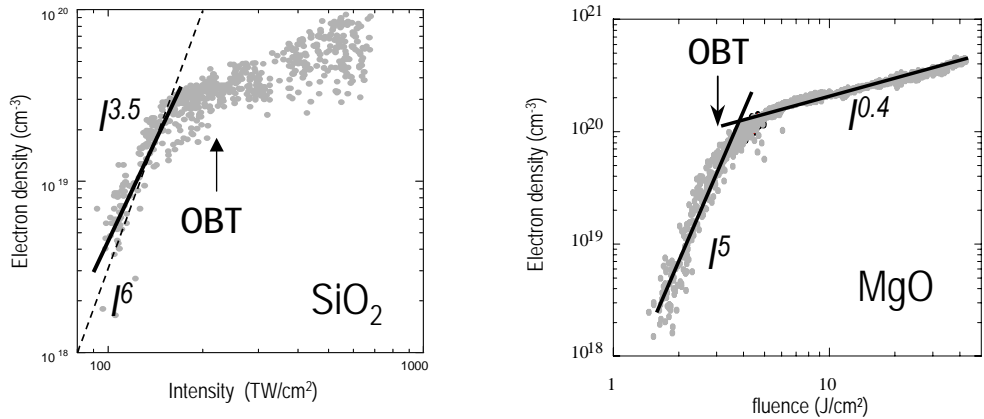

*Figure 18 : deux mesures de la densité de porteurs injectés par une impulsion titane/Saphir de 70 fs, au voisinage du seuil de claquage (OBT), pour $SiO_2$ et MgO. Attention à la différence des unités en abscisse !*

- **Thermique ou coulombien ?**

Cependant, si ces expériences apportent quelques éclaircissements sur les mécanismes d'excitation, elles ne nous disent pas l'origine du claquage : une forte densité de porteurs injectés signifie à la fois une fort transfert d'énergie de ceux ci au réseau, mais aussi une forte émission électronique de surface, qui peut induire des champs de charge d'espace importants en subsurface, et provoquer l'émission d'ions par un phénomène d'explosion coulombienne similaire à celui invoqué par Fleisher *et al.* (1975) dans le cas d'irradiations aux ions multichargés.

Le calcul d'Arnold et Cartier (1992), appliqué en particulier à des impulsions picosecondes d'un laser au Néodyme, simule la température du matériau, et celle ci dépasse la température de fusion de la silice au cours de l'impulsion, ce qui conduirait à la conclusion que là aussi le claquage est d'origine thermique. De quels indices expérimentaux disposons nous ?

Une expérience similaire à celle de la figure 18, répétée pour plusieurs durées d'impulsion sur MgO, ne montre pas de corrélation évidente entre seuil de claquage – monocoup - et densité de porteurs (Quéré, 2000) : la densité atteinte au moment du claquage décroît avec la durée de l'impulsion. Mais ceci semble toutefois insuffisant pour valider le modèle thermique.

Concernant l'explosion coulombienne, insistons sur le fait que c'est bien d'une émission collective ionique dont nous parlons. En effet le terme d'explosion coulombienne a souvent été utilisé dans le contexte de la physique moléculaire pour nommer le mécanisme par lequel une molécule à laquelle on enlève les électrons assurant la liaison va exploser sous l'influence de la répulsion coulombienne entre les ions qui la compose. Un mécanisme de ce type peut parfaitement exister dans les solides (Knotek et Feibelmann, 1978), mais il contribue plutôt à la désorption de surface qu'à l'ablation.

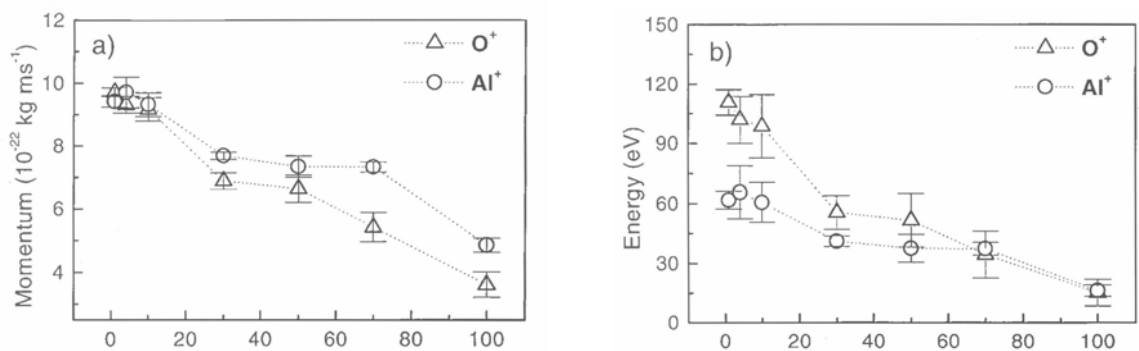

*Figure 19 : Moment (à gauche) et énergie cinétique (à droite) des ions émis par une surface d'$Al_2O_3$ sous irradiation par un laser titane-saphir (durée d'impulsion 100 fs), en fonction du nombre de tirs (Stoian et al., 2000)*

On peut chercher une méthode qui sonde l'état électrostatique de la couche de surface, et une idée consiste à utiliser pour cela les ions émis eux même. Ceux ci, à cause du champ induit en surface par le déficit local d'électrons, sont émis à partir d'un potentiel différent du potentiel moyen de l'échantillon, et cette distribution va se refléter dans la distribution d'énergie cinétique des ions émis. Dans le cas de l'explosion coulombienne, on s'attend à ce que tous les ions de même charge acquièrent le même moment. Au contraire, si le processus est d'origine thermique, c'est l'énergie cinétique des ions qui doit être la même, et déterminée par la température de l'échantillon.

Stoian *et al.* (2000) ont effectué une telle mesure sur Al$_2$O$_3$, en utilisant un laser titane-saphir, délivrant des impulsions de 100 fs dont le résultat est montré sur la figure 19. Il s'agit d'une expérience multicoups, et on constate que la situation évolue du début de l'expérience à la fin. Au début, on trouve bien que les ions Al$^+$ et O$^+$ ont le même moment, ce qui serait une indication forte en faveur du mécanisme d'explosion coulombienne. L'énergie des ions est élevée (60 eV pour Al, 120 eV pour O). Mais la situation évolue après un vingtaine de tirs, avec l'apparition d'une composante basse énergie, pour atteindre une situation (après 70 tirs) où c'est au contraire l'énergie cinétique des deux ions qui est égale (et faible : 10 eV). On constate en même temps que la morphologie de la surface ablatée évolue, montrant la transition d'une situation « d'ablation douce » vers une situation d'ablation « brutale », qui se manifeste aussi par une augmentation de l'émission lumineuse de la plume d'ablation, et une isotropisation de la distribution angulaire des ions émis.

La situation en début d'expérience serait celle qu'on rencontrerait dans une expérience monotir sur une surface peu perturbée. L'explosion coulombienne serait donc caractéristique d'une surface parfaite ou peu perturbée. Quant au caractère thermique de la distribution finale, il est difficile de dire si il résulte d'une évolution du processus d'ablation, ou de la thermalisation des ions au sein de la plume d'ablation. On a noté que, même pour des fluences très en dessous du seuil d'ablation, le rendement d'émission électronique d'un surface variait dramatiquement selon son état : il est pas exemple dans la silice multiplié par 10$^4$ quant on irradie une partie de la surface endommagée par un tir précédent. Il est donc évident que l'endommagement renforce le couplage laser-matériau, mais il est clair aussi qu'une surface endommagée comporte beaucoup plus d'électrons et d'ions faiblement liés.

Ce travail présente donc des éléments convaincants sur l'existence d'un processus « non thermique » dans l'ablation laser femtoseconde des isolants. Il n'est pas unique, et une expérience similaire sur le silicium a par exemple été publiée récemment (Roeterdink *et al.*, 2003), où on retrouve cet argument du moment comparé de deux espèces cette fois de même masse, mais de charge différente. Il est clair que plus de travail sur ce domaine est nécessaire pour clarifier la situation, mais que l'existence, dans certaines situations, d'un mécanisme non thermique dans l'ablation laser de matériaux isolants peut être considérée comme acquise.

- **Ablation en régime femto-, pico- et nanoseconde :**

Nous venons d'évoquer le fait que dans l'expérience de Stoian *et al.* (2000), il était difficile de décider si la thermalisation que l'on constate dans la distribution des ions après 70 tirs était le signe d'un changement de mécanisme d'ablation, ou si elle avait lieu plus tard, dans la phase d'expansion du plasma d'ablation. Nous arrivons là à un point décisif dans la description d'une expérience d'ablation : il ne suffit pas d'avoir identifié les mécanismes de base de l'interaction laser-solide, mais il faut aussi se préoccuper de ce qui se passe dans la phase d'expansion. Il est extrêmement difficile d'être exhaustif sur un tel sujet. Nous nous concentrerons donc sur quelques aspects essentiels, et sur quelques résultats explorant l'effet de la durée d'impulsion entre 70 fs et 10 ps, et sur une comparaison générale des régimes fs, ps et ns. Beaucoup des résultats mentionnés ici et non référencés sont tirés de la thèse, assez exhaustive de B. Sallé (1999).

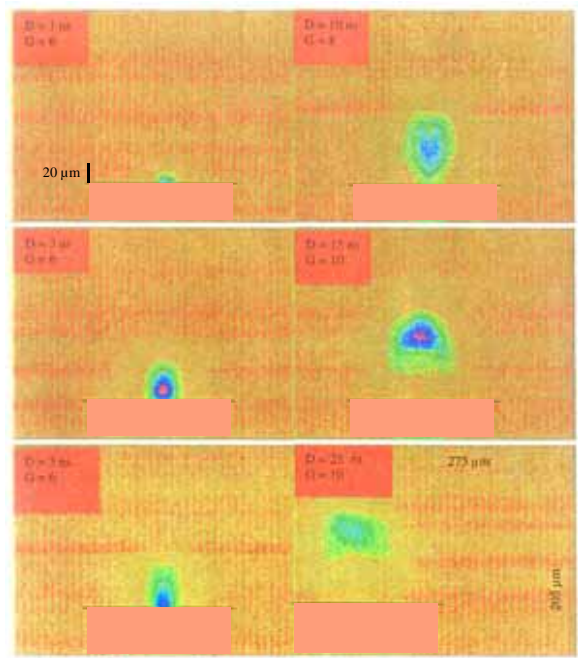

Figure 20 : images d'un plasma d'ablation du Cu pour une impulsion d'un laser titane-saphir de durée 800 fs, fluence : 15 J.cm$^{-2}$ (Sallé, 1999)

L'observation de l'évolution du plasma émis lors de l'ablation apporte de nombreux renseignements. La figure 20 présente des images du plasma d'ablation mesurées avec une caméra rapide, de 1 à 25 ns après l'interaction d'une impulsion de 70 fs d'énergie 20 µJ avec la surface d'un échantillon de cuivre (diamètre de la tache focale : 10 µm). On note un plasma bien défini, d'abord sphérique puis s'allongeant dans la direction perpendiculaire à la surface, de laquelle il se sépare très nettement. Ces images sont très différentes de celles qu'on obtiendrait (sauf à prendre des précautions particulières) avec des impulsions longues (présence de plusieurs points chauds, dont une zone chaude collée à la surface, forme oblongue, voir conique….) A partir des images de la figure 20 on peut tirer les ordres de grandeur suivants : on mesure des vitesses d'expansion longitudinale du plasma de $5\ 10^3$ m/s, et radiale de $3\ 10^3$ m/s (dans l'air). L'efficacité d'ablation est de $1.7\ 10^6$ µm$^3$/J, pour une impulsion de 20 µJ. On notera que les vitesses d'expansion sont proches de la vitesse du son dans le cuivre. Dans une expérience de ce type, on enlève par tir à la cible typiquement $0,5\ 10^{13}$ atomes. Le volume initial ablaté est de l'ordre de $10^{-10}$ cm$^3$ par tir, soit typiquement un profondeur ablatée de 10 nm. Evidemment, la mesure de l'expansion du plasma est faite plusieurs nanosecondes après l'interaction, et il serait hasardeux d'extrapoler le comportement quasi isotrope de l'expansion aux premiers instants, d'autant plus que dans sa phase initiale, cette expansion semble plus piquée vers l'avant (ce que confirme des expériences semblables – Garrelie *et al* (2001) – effectuées sous vide) . On fait donc l'hypothèse que l'expansion est 1D tant que l'expansion latérale n'excède pas 10% de la taille du faisceau, soit 1µm, ce qui correspond à une durée de 0.3 ns. A cet instant, l'extension longitudinale du plasma est de l'ordre de 1.5 µm, ce qui veut dire qu'on est passé d'une densité moyenne égale à celle du solide, à quelque chose de 150 fois plus faible. Par contre, en 70 fs (la durée de l'impulsion) l'extension calculée est de $3.5\ 10^{-10}$ m, complètement négligeable. Un autre chiffre intéressant est le temps qu'il faut pour abaisser la densité plasma à la densité critique à 800 nm : on trouve quelque chose de l'ordre de 12 ps. Il ne s'agit que d'ordres de grandeur, qui varieraient en fonction du matériau, et probablement aussi en fonction de l'environnement, mais ils peuvent servir de base à une typologie de l'interaction décrite dans le Tableau 1 ci dessous .

*Tableau 1*

*Types d'interaction laser-plasma-solide, en fonction de la durée d'impulsion laser*

| Durée de l'impulsion | <1 ps | 1ps à 10 ps | >10 ps |
|---|---|---|---|
| Type d'interaction | Laser-solide | Laser-plasma surcritique | Laser-plasma souscritique |
| Mode d'absorption | absorption intra/interbande dans l'épaisseur de peau du solide | absorption résonante à la densité critique | absorption par bremsstrahlung inverse dans le plasma |
|  | 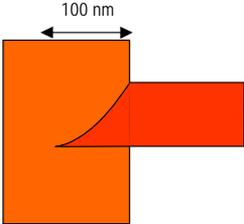 | 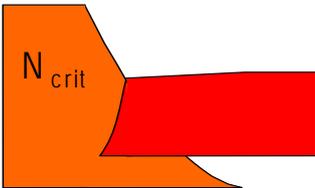 | 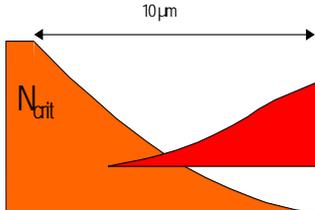 |

La différence essentielle entre les deux premiers types d'interaction est la suivante : c'est la « raideur » du gradient de densité. Dans le cas d'un solide, celle ci est pratiquement infinie, ce qui a pour conséquence que l' « absorption résonante » (qui se produit quand la densité du plasma est telle que l'énergie du plasmon est égale à celle du photon) se produit sur une épaisseur infiniment faible, et on peut donc la négliger. Dans un plasma, même surcritique (cas de la deuxième colonne) le gradient est beaucoup moins raide, et l'épaisseur sur laquelle cette absorption se produit est telle qu'elle va être le phénomène dominant. Dans un plasma sous critique, c'est l'absorption par bremsstrahlung inverse qui domine et chauffe de façon quasi homogène le plasma (dans lequel l'absorption est telle que l'intensité qui atteint la couche de densité critique est négligeable).

B. Sallé (1999) a étudié la puissance lumineuse totale émise par un tel plasma, pour plusieurs durées d'impulsions entre 70 fs et 10 ps , toutes les autres conditions expérimentales étant équivalentes à celles de la figure 20 (voir figure 21). On constate que celle ci baisse sensiblement entre 70 fs et 140 fs (probablement à cause de la disparition de contributions d'absorption multiphotoniques) puis augmente ensuite rapidement, ce qui est le signe de l'efficacité du mécanisme d'absorption résonante. Malheureusement, on ne dispose pour traiter ce type de situation (couplant l'évolution hydrodynamique du matériau, et les phénomènes radiatifs), que des codes assez lourds utilisés dans le cadre des études à très haute intensité (fusion contrôlée par laser). Mais ces codes (qui sont déjà en questions quand on approche de situations « hors équilibre thermodynamique local ») ne s'appliquent pas vraiment aux situations de flux modérés qui sont notre sujet ici, et pour lesquelles plusieurs approximations utilisées sont à revoir. Néanmoins, ce qui est exposé ci dessus permet d'énoncer la règle suivante :

*L'utilisation d'impulsions femtoseconde permet de considérer qu'on interagit avec un matériau dont la détente hydrodynamique n'a pas commencé. On est donc bien dans ce cas – et dans celui là seulement - en face d'une interaction laser-solide*

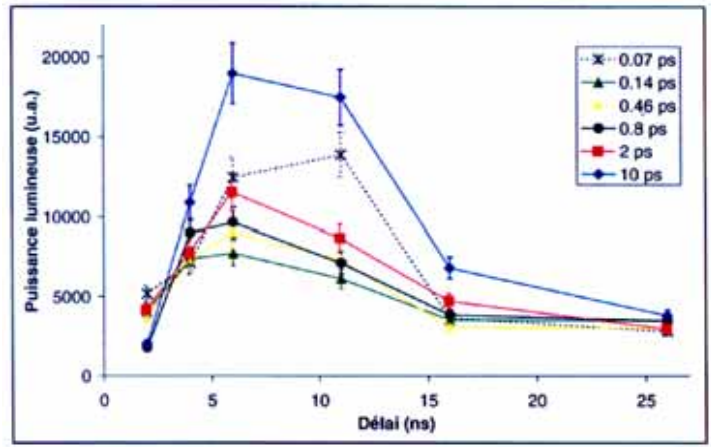

*Figure 21 : Emission lumineuse du plasma de la figure 20, pour différentes durées d'impulsions (fluence constante)*

Nous n'avons pas montré ici les images comparant des cratères d'ablation obtenus en impulsion nanoseconde et picoseconde. Elles sont nombreuses et montrent la même chose : le cratère est plus propre en impulsion femtoseconde, et cela a d'abord été attribué à l'absence d'effets thermiques. Nous ne le croyons pas et pensons au contraire que ceci s'explique par la différence fondamentale des interactions mises en jeu :
- en impulsion femtoseconde, on a vraiment affaire à une interaction laser-solide
- en impulsion nanoseconde, le laser interagit pour l'essentiel avec le plasma d'ablation, et c'est ce dernier qui interagit avec la surface. Les fortes pressions qu'il exerce sur un matériau fondu sont la meilleure explication aux effets de bord constatés dans ce cas.
- en impulsion picoseconde, des situations intermédiaires peuvent être obtenues, comme par exemple des cratères de formes complexes (un cratère profond au centre, entouré d'une dépression large et peu profonde). Ceci s'explique grâce à la réfraction du faisceau sur le plasma en expansion, dont la densité est encore suffisante pour avoir un effet défocalisant important.

Il est intéressant de noter que des techniques de focalisation spécifiques ont été utilisées pour obtenir un découplage maximum du plasma (au delà évidemment de l'effet connu de la longueur d'onde : il faut travailler dans l'UV). L'une d'entre elle consiste à utiliser un objectif de grande ouverture numérique (type Cassegrain) de façon à faire arriver un maximum de lumière par les cotés du plasma. Semerok *et al.* (1999) ont utilisé cette technique, et comparé l'efficacité d'ablation de lasers nano, pico et femtoseconde (ces deux derniers étant focalisés normalement). Les résultats de cette comparaison sont montrés sur le tableau 2

*Tableau 2*

*Efficacité d'ablation en $10^8$ µm³/J, énergie de l'impulsion : 60 µJ*

| Cible | Laser nanoseconde Nd-YAG | | Laser picoseconde nd-YAG | | | laser femtoseconde Ti-$Al_2O_3$ |
|---|---|---|---|---|---|---|
| | $\lambda$ =532 nm $d$ =8µm | $\lambda$ =266 nm $d$ =8µm | $\lambda$ =1064 nm $d$ =20µm | $\lambda$ =532 nm $d$ =9.5µm | $\lambda$ =266 nm $d$ =6.5 µm | $\lambda$ =400 nm $d$ =16µm |
| Al | .0124 | 0.293 | 0.010 | 0.040 | 0.057 | 0.204 |
| Cu | 0.031 | 0.065 | 0.004 | 0.009 | 0.028 | 0.050 |

Notons que si on utilise une optique normale pour le laser nanoseconde, on retrouve des chiffres (à longueur d'onde équivalente) qui se rapprochent des chiffres obtenus en picoseconde. On constate bien que ce sont les situations où on a « débranché » l'interaction laser-plasma d'ablation qui se révèlent les plus efficaces (la faible différence femtoseconde / nanoseconde s'expliquant par une effet de longueur d'onde).

# Conclusion

Nous avons rappelé les processus fondamentaux qui interviennent dans l'interaction laser-solide, qui constituent les briques élémentaires d'une expérience d'ablation laser. Ces processus décrivent les mécanismes d'absorption de l'énergie laser par les électrons du matériau, et comment cette énergie est répartie entre les électrons du matériau, et finalement transmise au réseau cristallin, sous forme de température, ou d'énergie associée à la polarisation du réseau cristallin, indissociable du piégeage des porteurs.

Nous avons décrit quelques grands modèles intégrant ces processus élémentaires en une description d'ensemble de l'interaction. Pour les métaux, malgré ses imperfections fondamentales, le modèle « thermique » - à deux températures - semble au moins fonctionner. Le cas des isolants est à l'évidence plus difficile, et il est clair qu'on est loin d'avoir atteint un consensus sur quels phénomènes dominent pour expliquer par exemple le claquage optique en impulsions courtes et intenses. De même, si il semble clair que des mécanismes de type « explosion coulombienne » existent dans les isolants, on est loin d'avoir déterminé leur domaine de prévalence vis à vis des phénomènes thermiques qui peuvent aussi exister ici. Nous avons par contre peu commenté, ce qui aurait été trop long, les descriptions assez détaillées des phénomènes de création de défauts ponctuels dans les isolants, dont on dispose notamment grâce aux expériences de type pompe-sonde en impulsions femtoseconde. Ils sont cependant importants car ils peuvent constituer la base d'un « vieillissement optique » du matériau dans des expériences multicoups qui correspondent à la plupart des applications pratiques

Nous avons identifiés un certain nombre de caractéristiques particulières des impulsions femtoseconde utilisées dans un tel contexte. On peut les résumer ainsi :
- l'utilisation d'impulsions femtoseconde renforce, dans les matériaux isolants, les processus intrinsèques par rapport à ceux qui résultent de la présence de défauts. Cette propriété résulte notamment du fait qu'il est possible avec des impulsions plus courtes de travailler à des éclairements plus élevés sans atteindre le seuil de dommage du matériau.
- l'utilisation d'impulsions femtoseconde permet de considérer que l'interaction se produit avec un matériau dont les vibrations ont été gelées.
- l'utilisation d'impulsions femtoseconde peut donner lieu à des situations de déséquilibre thermodynamique notable entre les électrons du matériau et le réseau. Ces situations sont naturellement transitoires, mais peuvent avoir des conséquences physiques importantes (par exemple sur l'émission électronique de la surface irradiée).
- si l'utilisation d'impulsions femtoseconde permet effectivement de révéler l'existence de phénomènes « non thermiques », il n'est pas vrai pour autant qu'elle supprime ces derniers .
- l'utilisation d'impulsions femtoseconde permet de découpler la phase d'interaction laser-matière de la phase d'expansion thermodynamique du matériau irradié.

Cette dernière propriété nous semble la plus importante d'un point de vue pratique. On peut affirmer que c'est seulement dans le cas d'impulsions femtoseconde, dans une expérience d'ablation laser, qu'on a affaire réellement à une interaction laser-solide. C'est très probablement l'explication principale des différences notables des profils des cratères obtenus en ablation nanoseconde et femtoseconde. De ce point de vue, l'utilisation d'impulsions femtoseconde induit une situation plus simple du point de vue de la physique de l'interaction, ce qui ne veut pas forcément dire qu'elles soient toujours intéressantes d'un point de vue pratique. Il est clair par contre qu'elles offrent beaucoup d'avantages pour ce qui est de la recherche détaillée sur les mécanismes de l'ablation laser, un domaine sur lequel il reste beaucoup à faire.

Parmi les directions à explorer, on retiendra sûrement l'étude du phénomène d'explosion coulombienne, et des mécanismes « non-thermiques » qui interviennent sûrement dans le cas des diélectriques (bien que les aspects thermiques ne doivent pas être oubliés) et qui sont encore mal connus. L'étude de la dynamique des plasmas d'ablation est aussi prometteuse, particulièrement si on lui applique les techniques pompe-sonde, qui permettraient d'obtenir plus d'information sur les phases initiales de l'expansion , inaccessibles par des méthodes classiques (caméras image-par-image, mêmes ultrarapides), et qui devraient apporter plus d'information sur l'état du matériau en fin d'impulsion. Particulièrement intéressantes de ce point de vue sont les technique d'imagerie interférométriques : elles ont donné d'excellents résultats avec des impulsions picosecondes pour le cas de plasma dans l'air ou de plasmas électrons-trous dans les isolants (Garnov *et al.*, 2003). Enfin la théorie mérite certainement un sérieux effort, particulièrement dans le cas des diélectriques. Il s'agit évidemment de résoudre les problèmes encore en débat qui ont été signalés plus haut, avec le souci d'avancer vers des modèles simplifiés, retenant ce qui est essentiel, et capables de produire des prédictions fiables, tout en étant suffisamment simples pour être extrapolables des matériaux idéaux étudiés jusqu'ici aux systèmes d'intérêt pratique, nécessairement plus complexes.

# Bibliographie